\documentclass[12pt]{article}

\begin{document}

\author{A.I.Volokitin$^{1,2}$ and B.N.J.Persson$^1$ \\
\\
$^1$Institut f\"ur Festk\"orperforschung, Forschungszentrum \\
J\"ulich, D-52425, Germany\\
$^2$Samara State Technical University, 443100 Samara,\\
Russia}
\title{Resonant photon tunneling enhancement of the radiative heat transfer}
\maketitle

\begin{abstract}
We study the dependence of  the heat transfer  between two semi-infinite 
solids  on the dielectric properties of the bodies.  
We show that  the  
heat transfer at short separation between the solids may increase  by many order of
magnitude when the surfaces are covered by adsorbates, or can support
low-frequency surface plasmons. In this case
the heat transfer  is determined by resonant photon tunneling between adsorbate
vibrational modes, or surface plasmon modes.
We study the dependence of  the heat flux between  two metal surfaces  on 
the electron concentration using the non-local optic dielectric approach, and compare 
 with the results obtained within local optic approximation.
\end{abstract}

\section{ Introduction}

It is well known that for bodies separated by $d>>d_T= c\hbar /k_BT$, the
radiative heat transfer between them is described by the Stefan- Bolzman
law:
\begin{equation}
S=\frac{\pi ^2k_B^4}{60\hbar ^3c^2}\left( T_1^4-T_2^4\right)
, \label{stefan}
\end{equation}
where $T_1$ and $T_2$ are the temperatures of solid $1$ and $2$,
respectively. In this limiting case the heat transfer is connected with
traveling electromagnetic waves radiated by the bodies, and does not depend
on the separation $d$. For $d<d_T$ the heat transfer increases by many order
of magnitude due to the  evanescent
 electromagnetic waves  that decay exponentially into the vacuum; this is offen refered to as 
photon tunneling.  At present
 there is an increasing number of investigations of heat transfer due to
evanescent waves in connection with the scanning tunneling microscopy, and
the scanning thermal microscopy (STM) under ultrahigh vacuum conditions \cite
{Van Hove,Levin1,Pendry,Majumdar,Volokitin2,Volokitin3,Mulet}. 
 It is now possible to measure extremely small amounts of heat transfer into small
 volumes \cite{Barnes}. STM can be used for local heating of the
surface, resulting in local desorption or decomposition of molecular
species, and this offer further possibilities for the STM to control local
chemistry on a surface.

The efficiency of the radiative heat transfer  depends strongly on the dielectric 
properties of the media. In \cite{Pendry,Volokitin2,Volokitin3} it was shown  
the heat flux can be greatly enhanced if the conductivities of the material 
is chosen to maximize the heat flow due to photon tunneling. At room temperature 
the heat flow is maximal at conductivities corresponding to semi-metals. In fact, 
only a  thin film ($\sim10${\AA}) of a  high-resistivity material  is needed to maximize 
the heat flux. Another enhancement mechanism of the radiative heat transfer can 
be connected with resonant photon tunneling between states localized on the 
different surface. Recently it was discovered that resonant photon tunneling between
surface plasmon modes give rise to extraordinary enhancement of the optical
transmission through sub-wavelength hole arrays \cite{Ebbesen}. The same
surface modes enhancement  can be expected  for the radiative heat transfer (and the 
van der Waals friction \cite{Volokitin}) if the
frequency of these modes is sufficiently low to be excited by thermal radiation.
At room temperature only the modes with frequencies below
$\sim 10^{13}s^{-1}$ can be excited.
For normal metals surface plasmons have much too high frequencies; at
thermal frequencies
the dielectric function of normal metals becomes nearly
purely imaginary, which exclude surface plasmon enhancement of the heat 
transfer  for good conductors. However surface plasmons
for semiconductors are characterized by much
smaller frequencies and damping constants, and they can give an important contribution
to the heat transfer. Recently,  enhancement of the heat 
transfer due to resonant photon tunneling between surface plasmon modes localized on 
the surfaces of the semiconductors was predicted in \cite{Mulet}. The authors studied the 
radiative heat transfer between a small particle, considered as a point-like dipole, and 
a flat surface. However, this treatment can be applied for scanning probe microscopy only 
in the case $R>>d$, where $R$ and $d$ are the radius of the particle and the separation 
between the particle and the surface, respectively.  For the opposite limit $R<<d$, which is more appropriate 
for the scanning probe microscopy, the heat transfer between the tip and the surface can, in the 
first approximation,  be modeled by the heat transfer between two semi-infinite solids. In 
 this case the multiple scattering of the electromagnetic waves by the surfaces of the bodies, 
which was not taking into account in \cite{Mulet}, becomes important in the photon tunneling. 
In particular, at sufficiently small separation $d$, the photons goes back and forth several time 
in the vacuum gap, building up coherent constructive interference in the forward direction 
much as would occur in resonant electron  tunneling. In this case the surface plasmons on the isolated 
surface combine to form a "surface plasmon molecule", in much the same way as electronic states 
of isolated atoms combine to form molecular levels. This will result in a  very weak distance 
dependence of the heat flux, because the transmission probability for photon does  not depend on $d$ 
in this case 
(see below). For large $d$ the  sequential tunneling is more likely to occur, where the  
photon excited in a surface plasmon mode, tunnels to the surface plasmon at the other surface, and then 
couples to the other excitations in the media and exit.   
Other  surface modes which can be excited
by thermal radiation are
adsorbate vibrational modes. Especially for parallel vibrations these modes
may have very low frequencies.

All information about the long-range electromagnetic interaction between two
non-contacting bodies is contained in the reflection factors
of the electromagnetic field. At present time very little is known about the
reflection factors for large wave vectors and for extremely small
frequencies.  
In our previous calculations of the radiative heat transfer and Van der Waals friction \cite
{Volokitin1,Persson1,Volokitin2,Volokitin3} we mostly
considered good conductors. In this case it was shown that the important
contribution comes from the non-local optic effects in the surface region.
However it was shown that the radiative heat transfer and Van der Waals friction becomes much larger for
high resistivity material, for which the volume contribution from non-local
effects is also important. Non-local optic refer to the fact that the current 
at point $\mathbf{r}$ depends on the electric field not only at point $\mathbf{r}$, 
as it is assumed in the  local optic approximation, but also at points $\mathbf{r}^{\prime}
\neq\mathbf{r}$ in a finite region around the point $\mathbf{r}$. In the case when both points 
are located outside the surface region the dielectric response function can be expressed through the dielectric 
function appropriate for the semi-infinite electron gas. However, if one of the point $\mathbf{r}$ 
or $\mathbf{r}^{\prime}$ is located in the surface region,  the dielectric response function will be 
different from its volume value, and this gives the surface contribution from nonlocality. 
 In order to verify the accuracy of the local optic approximation we study the
dependence of the radiative heat transfer on the dielectric properties of the
materials within the non-local dielectric approach, which was proposed some
years ago for the investigation of the anomalous skin effects \cite{Fuchs
and Kliever}. 

\section{Theory}

The problem of the radiative heat transfer between two flat surfaces  was considered some years
ago by Polder and Van Hove \cite{Van Hove}, Levin and Rytov \cite{Levin1}
and more recently by Pendry \cite{Pendry}, and Volokitin and Persson \cite{Volokitin2,Volokitin3}. 
Polder and Van Hove were the first who obtained the correct formula for the heat transfer
 between two flat surface. In their investigation  they used  Rytov's theory \cite{Rytov,Levin2} of the 
fluctuating electromagnetic field. However, they presented their  result 
only for identical media, and  within the local optic 
approximation. In the subsequent treatment, they made numerical calculations  
 not of the heat flux itself, but its derivative with respect to temperature,  
 i.e., their result pertain only to the small temperature differences. Unfortunately, 
their paper contains no analytical formulas in closed form . Levin and Rytov 
\cite{Levin1} used the generalized Kirchhoff's law \cite{Levin2} to obtain  
 an expression for the radiative heat transfer between two semi-infinite media 
in the impedance approximation. The case of the good conductors was investigated
in the details both in   normal and the anomalous skin effect region. Pendry \cite{Pendry} 
proposed a  more compact derivation of the formula for the heat flux between two semi-
infinite bodies due to evanescent waves, and calculated the heat transfer between a point- 
dipole and a surface.  
 Volokitin and Persson \cite{Volokitin2} considered the problem of the heat transfer between two flat surfaces, 
 as a particular case of the general approach 
for calculation of the heat transfer. They investigated numerically the  dependence of the heat 
flux   on the dielectric properties of the bodies, and 
found that for good conductors, even for very small distances the heat flux is dominated by 
 retardation effects. They also showed that the heat flux heat is maximal at conductivities, 
corresponding to the semi-metal. 

According to \cite{Van Hove,Levin1,Pendry,Volokitin2}   
 the heat transfer between two semi-infinite bodies , separated by a vacuum gap
with the width $d$, is given by the formula
\begin{equation}
S=\int_0^{\infty}d\omega\left(\Pi_1-\Pi_2\right)M
\end{equation}
where
\[
M=\frac 1 {4\pi ^2}
\int_0^{\omega/c} dq\,q
 \frac{(1-\mid R_{1p}(\omega )\mid ^2)(1-\mid R_{2p}(\omega
)\mid ^2)}{\mid 1-\mathrm{e}^{2i
pd}R_{1p}(\omega )R_{2p}(\omega )\mid ^2}  
\]
\begin{equation}
+\frac{1}{\pi^2}\int_{\omega/c}^{\infty} dq\,q\mathrm{e}^{-2k d}  
\times \frac{\mathrm{Im}R_{1p}(\omega )\mathrm{Im}R_{2p}(\omega )}{\mid 1-%
\mathrm{e}^{-2\mid p\mid d}R_{1p}(\omega )R_{2p}(\omega )\mid ^2}
 +\left[ p\rightarrow s\right]   \label{one}
\end{equation}
where the symbol $\left[ p\rightarrow s\right]$ stands for the terms 
which can are obtained from the first two terms by replacing the reflection factor $R_p$ for the 
$p$-polarized electromagnetic waves  with the reflection factor $R_s$ for $s$- 
polarized electromagnetic waves ,
and where $p=((\omega/c)^2-q^2)^{1/2},\, k=|p|$. The Plank function of solid \textbf{1}  
\begin{equation}
\Pi_1(\omega )=\hbar \omega \left( e^{\hbar \omega /k_BT_1}-1\right) ^{-1},
\end{equation}
 and similar for $\Pi_2$. The contributions to the heat transfer 
from the propagating ($q<\omega/c$) and evanescent ($q>\omega/c$) electromagnetic waves are determined 
by the first and the second terms in Eq. (\ref{one}), respectively.

Let us firstly consider some general  consequences of Eq. (\ref{one}).
In the case oh heat transfer through free photons  ($q \le \omega/c$), the transfer is maximal 
when both bodies are perfectly black and have zero reflection coefficient, $R=R_r + iR_i=0$.
Now, what is the photon- tunneling equivalent of a black body? For $q>\omega/c$ there are no constraints 
on the reflection coefficient $R(q,\omega)$ other than that $\mathrm{Im}R(q,\omega)$ is positive. 
Therefore, assuming identical surfaces, we are free to maximize the transmission coefficient corresponding 
to the photon tunneling 
\begin{equation}
T=\frac{R_i^2\mathrm{e}^{-2kd}}{\left|1-\mathrm{e}^{-2kd}R^2\right|^2}
\label{hten}
\end{equation}
This function is a maximum when \cite{Pendry}
\begin{equation}
R_r^2+R_i^2=\mathrm{e}^{2kd}
\end{equation}
so that $T=1/4$. Substituting these result in (\ref{one}) gives the evanescent contribution 
\begin{equation}
(S_z)_{max}^{evan}=\frac{k_B^2T^2q_c^2}{24\hbar}
\label{heleven}
\end{equation}
where $q_c$ is a cut-off in $q$, determined  by the properties of the material. 
 It is clear that the 
largest possible  $q_c \sim 1/a$, 
where $a$ is an interatomic distance. Thus, from Eq. (\ref{heleven}) we get upper 
boundary for the radiative heat transfer at  room temperature: 
$(S_z)_{max} \sim 10^{12}\mathrm{Wm^{-2}}$.

We rewrite the denominator of the integrand in the term in Eq. (\ref{one}), which  
corresponding to the evanescent waves, in the form
\[
|1-e^{-2kd}R|^2=[(1-e^{-kd}R_r)^2+e^{-2kd}R_i^2]
\]
\begin{equation}
\times [(1+e^{-kd}R_r)^2+e^{-2kd}R_i^2]
\label{5one}
\end{equation}
The conditions for resonant photon tunneling are determined by equation
\begin{equation}
R_r(\omega_{\pm}(q))=\pm e^{kd}
\label{two}
\end{equation}   
This  condition can be fulfilled even 
when $\mathrm{exp}(-2kd)<<1$ because for evanescent electromagnetic waves there is no
restriction on the magnitude of real part or the modulus of $R$.
This open up the possibility of
resonant denominators for $R_r^2>>1$.
Close to resonance we can write
\[ 
(1\pm e^{-kd}R_r)^2\pm e^{-2kd}R_i^2
\]
\begin{equation}
\approx e^{-2kd} R_r^{\prime 2}(\omega_{\pm})[(\omega-\omega_{\pm})^2+
(R_i(\omega_{\pm})/ R_r^{\prime}(\omega_{\pm}))^2],
\label{three}
\end{equation}
where
\[
R_r^{\prime}(\omega_{\pm})=\left.\frac{dR_r^{\prime}(\omega)}{d\omega}
\right|_{\omega=\omega_{\pm}},
\]
which leads to the following contribution to the heat flux:
\begin{equation}
S_{\pm} \approx \frac{1}{4\pi}\int_0^{q_c}dk\,k
\left(\Pi_1(\omega_{\pm})-\Pi_2(\omega_{\pm})\right)
\frac{R_i(\omega_{\pm})}{R_r^{\prime}(\omega_{\pm})}.    
\label{four}
\end{equation}
The parameter $q_c$ in this expression defines the region  $0<q<q_c$ where 
the two- poles approximation is valid. To proceed further, let us make 
the following simplifications. Close to a  pole we can  
 use the approximation
\begin{equation}
R=\frac{a}{\omega-\omega_0-i\eta},
\label{six}
\end{equation}  
where $a$ is a constant. Then from the resonant condition (\ref{two}) we get
\[ 
\omega_{\pm}=\omega_0\pm ae^{-kd}.
\]
For the two poles approximation to be valid the difference $\Delta\omega =
|\omega_+-\omega_-|$ must be greater than the width $\eta$ of the resonance. 
 From this condition we get
$q_c\le\ln(2a/\eta)/d$.
For 
short distances the parameter $q_c$ defines the value of $q$ where 
the solution of Eq. (\ref{two}) ceases to exist. 

For $\omega_0>a$ and $q_cd>1$, from Eq. (\ref{four}) we get
\begin{equation}
J_{\pm}=\frac{\eta q_c^2}{8\pi}\left[\Pi_1(\omega_0)-\Pi_2(\omega_0)\right].
\label{seven}
\end{equation}

Interesting, the explicit $d$ dependence has dropped out of Eq. (\ref{seven}). 
However, $J$ may still be $d$- dependent, through the $d$- dependence 
of $q_c$. For the small distances one can expect that $q_c$ is determined 
by the dielectric properties of the material, and thus does not depend on $d$. In this 
case the heat transfer  will be also distance independent.  
           
\section{Numerical results}

Resonant photon tunneling enhancement of the heat transfer 
 is possible for two semiconductor surfaces which can support
low-frequency surface plasmon modes. The reflection factor $R_{p}$
for clean semiconductor surface  is given by Fresnel's formula 
\begin{equation}
R_{p}=\frac {k-s/\epsilon}
{k+s/\epsilon}, \label{eight}
\end{equation}
where
\begin{equation}
s=\sqrt{k^2-\left(\frac{\omega}{c}\right)^2(\epsilon-1)},   \\
\end{equation}
where $\epsilon$ is the bulk dielectric function.
As an example, we consider
two clean surfaces of silicon carbide (SiC). The
optical properties of this material can be described using an oscillator
model \cite{Palik}
\begin{equation}
\epsilon(\omega)=\epsilon_{\infty}\left(1+\frac{\omega_L^2 - \omega_T^2}
{\omega_T^2 - \omega^2 -i\Gamma \omega}\right)
\label{nine}
\end{equation}
with $\epsilon_{\infty}=6.7, \,\omega_L=1.8\cdot10^{14}s^{-1},\,
\omega_T=1.49\cdot10^{14}s^{-1},\,$ and $\Gamma=8.9\cdot10^{11}s^{-1}.$ The
frequency of surface plasmons is determined by condition
$\epsilon_r(\omega_p)=
-1$ and from (\ref{nine}) we get $\omega_p=1.78\cdot10^{14}s^{-1}$. In Fig.1
we plot the heat flux  $S(d)$: note that the
heat flux  between the two semiconductor surfaces is  several order of magnitude
larger than between two clean  good conductor surfaces (see Fig.3).

Another enhancement mechanism is connected with resonant photon tunneling
between adsorbate vibrational modes localized on different surfaces. 
As an example, let us consider ions with charge $e^*$ adsorbed on metal surfaces.
The reflection factor $R_{p}$, which
takes into account the contribution from an adsorbate layer, is given by \cite{Langreth}:
\begin{equation}
R_{p}=\frac {q-s/\epsilon+4\pi n_aq[s\alpha_{\parallel}/\epsilon+q \alpha_{\perp}]}
{q+s/\epsilon+4\pi n_aq[s\alpha_{\parallel}/\epsilon-q \alpha_{\perp}]}, \label{ten}
\end{equation}
and where $\alpha_{\parallel}$ and $\alpha_{\perp}$ are the
polarizabilities of adsorbates
in a direction parallel and normal to the surface,  respectively. $\epsilon$
is the bulk  dielectric function and $n_a$ is
the concentration of adsorbates. For clean surfaces $n_a=0$, and in this
case formula (\ref{ten}) reduces to the  Fresnel formula.
The polarizability for ion vibration normal to the surface is given by
\begin{equation}
\alpha_{\perp}=\frac {e^{*2}}{M(\omega_{\perp}^2-\omega ^2 -i\omega \eta_{\perp})},
\end{equation}
where $\omega_{\perp}$ is the frequency of the normal adsorbate  vibration,
and $\eta_{\perp}$ is the damping constant.
In Eq. (\ref{ten}) the contribution  from parallel vibrations
is reduced by the small factor $1/\epsilon$. However, the contribution
of parallel vibrations to the heat transfer
 can nevertheless   be important
due to the indirect interaction of the parallel adsorbate vibration with the electric field,
via the metal conduction electron \cite{Persson and Volokitin2}. Thus, the small
parallel component
of the electric field will induce a strong electric current in the metal.
The drag force between
the electron flow and the adsorbates can induce adsorbate vibrations  parallel to the
surface. This gives the polarizability:
\begin{equation}
\alpha_{\parallel}=\frac {\epsilon -1}{n}\frac {e^{*}}{e}\frac {\omega \eta_{\parallel}}
{(\omega^2_{\parallel}-\omega^2 -i\omega \eta_{\parallel})}
\end{equation}
where $n$ is the conduction electron concentration.
As an illustration, in Fig.2 we show the heat flux  for the two
Cu(001) surfaces covered by a low concentration of potassium atoms
( $n_a=10^{18}m^{-2}$)
. In  the $q-$
integral in Eq.(\ref{one}) we used the cut off  
$q_c \sim \pi/a$ (where $a\approx1nm$
is the inter-adsorbate
distance) because our microscopic approach is applicable only when the wave length
of the electromagnetic field is larger than double average distance between the
adsorbates.
 In comparison, the heat flux
between two clean surface at separation $d=1nm$ is two order of
magnitude smaller.

Fig.3 shows the thermal flux between two clean metal surfaces as a
function of electron density $n.$ In the calculations we have assumed that
one body is at zero temperature and the other at $T=273\,\,$K, and the Drude
relaxation time $\tau =4\cdot 10^{-14}$s$^{-1}.$ When the electron density 
decreases there is transition from a degenerate electron gas ($k_BT<<\varepsilon_F$, 
where $\varepsilon_F$ is the Fermi energy) to a non-degenerate electron gas ($k_BT>>\varepsilon_F$) 
at the density $n_F\sim(K_BTm)^{3/2}/\pi^2\hbar^3$, where $m$ is the electron mass. At $T=273\,\,$K 
the transition density $n_F\sim$10$^{25}$m$^{-3}$. 
The full line was obtained
by interpolation between the two dashed curves, calculated in the non-local
dielectric function formalism for the non-degenerate electron gas (valid for
$n<n_F\approx $ $10^{25}$m$^{-3})$, and for the degenerate electron gas (for $%
n>n_F)$ \cite{Fuchs and Kliever}. The thermal flux reaches the maximum $S_{\max }\approx
5\times 10^8$W$\cdot $m$^{-2}$ at $n_{\max }\approx 10^{25}$m$^{-3},$ which
corresponds to the DC conductivity $\sigma \approx 3\cdot 10^3$($\Omega
\cdot $m)$^{-1}.$ 
Within the local optic approximation the  radiative heat
transfer is maximal at $n_{L\max }\approx 10^{24}$m$^{-3}$ where $S_{L\max
}\approx 10^9$W$\cdot $m$^{-2}$. The thermal flux due to traveling
electromagnetic waves is determined by formula (\ref{stefan}) which gives $%
S_{BB}=308\,$W$\cdot $m$^{-2}$ for $\,T=273\,K.$

\section{Summary}

 We have studied the radiative heat transfer in dependence on the 
dielectric properties of the media. We have found that at sufficiently  
short distances between bodies the thermal flux can be significantly
enhanced in comparison with the black body radiation 
 when the material involved support low-frequency adsorbate vibrational 
modes or surface plasmon modes, or the conductivity of the metals 
is chosen to optimize the heat transfer. 
This fact can be used in scanning probe microscopy for local heating and
modification of surfaces.

\vskip 0.5cm \textbf{Acknowledgment }

A.I.V acknowledges financial support from DFG 
, B.N.J.P.  acknowledges 
support from the European Union Smart Quasicrystals project.

\thinspace \thinspace \thinspace

\vskip 0.5cm

FIGURE CAPTIONS

Fig. 1. The heat flux between  two semiconductor surfaces 
 as a function of
separation $d$ . One body is at zero temperature and the other 
at $T=273\,$K For parameters  corresponding to a surface of silicon 
carbide (SiC) (see text for the explanation). 
 (The log-function is with
basis 10)

Fig. 2. The heat flux between two surfaces covered by adsorbates 
 , as a function of the
separation $d$. One body is at zero temperature and the other
at $T=273\,$K. For parameters   corresponding to K/Cu(001)
\cite{Senet} ($ \omega_{\perp}=1.9\cdot 10^{13}s^{-1}, \omega_{\parallel}=
4.5\cdot 10^{12}s^{-1}, \eta_{\parallel}=2.8\cdot 10^{10}s^{-1},
 \eta_{\perp}=1.6\cdot 10^{12}s^{-1}, e^{*}=0.88e$)
(The log-function is with basis 10)

Fig. 3. The heat flux between two metal surfaces as a function of the free
electron concentration $n$. One body is at zero temperature and the other at
$T=273\,\,K.$ The full line was obtained by interpolation between curves
(dashed lines) calculated in the non-local dielectric formalism 
 for a non-degenerate electron gas for $%
n<n_F\sim $ $10^{25}m^{-3}$ , and for a degenerate electron gas for $n>n_F$.
Also shown is results (dashed lines) obtained within the local optic
approximation. The calculation were performed with the damping constant $%
\tau ^{-1}=2.5\cdot 10^{13}s^{-1}$ , separation $d=10$\AA\ and $n_0=8.6\cdot
10^{28}m^{-3}.$ (The log-function is with basis 10)

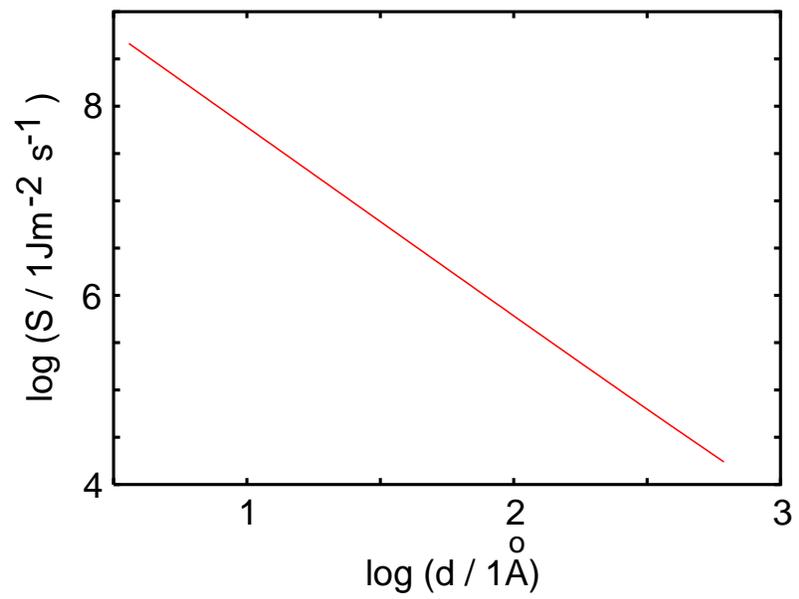
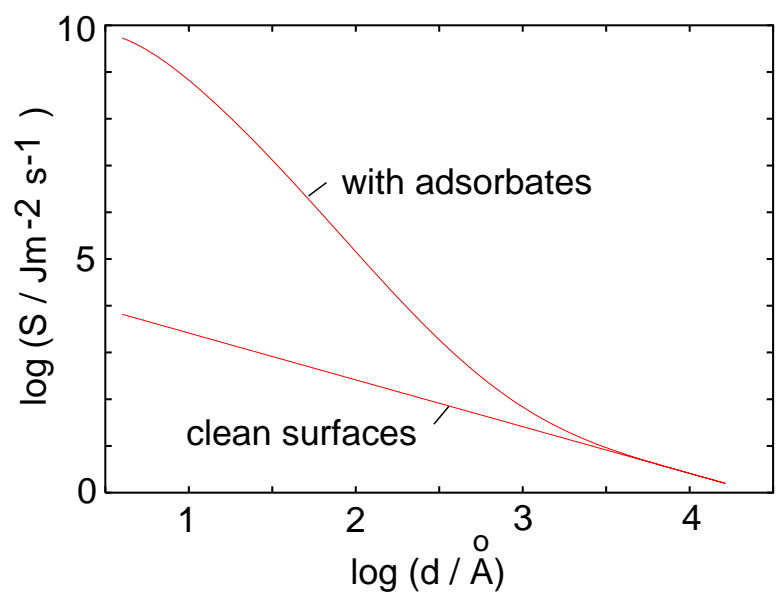
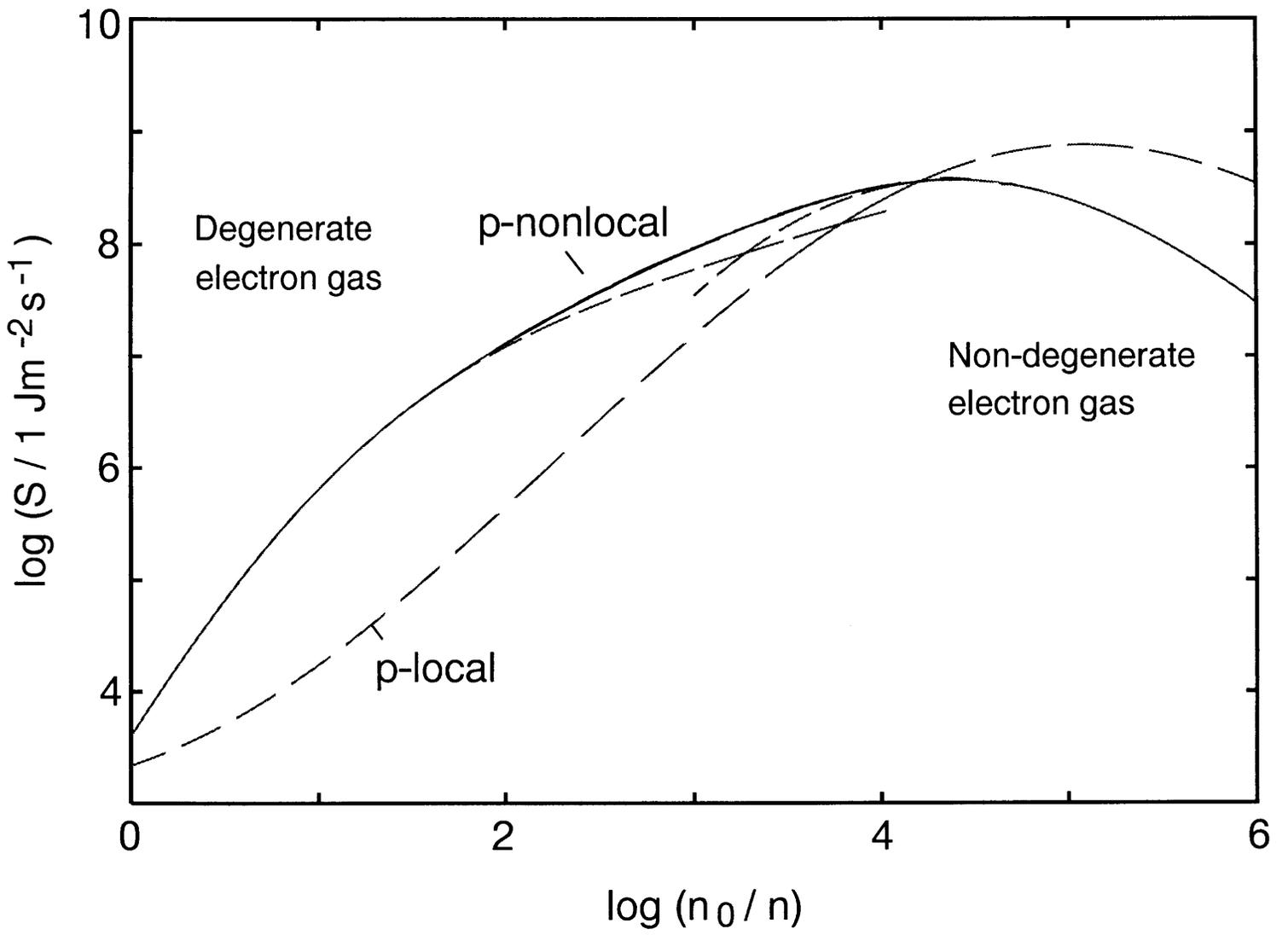
\end{document}